\documentclass[aps,prl,reprint,nofootinbib, preprintnumbers, superscriptaddress]{revtex4-1}

\usepackage{bm}
\usepackage{graphicx} %
\usepackage{amsmath} %
\usepackage{amssymb} %
\usepackage{url}
\usepackage{subfigure} %
\usepackage{float} %
\usepackage{upgreek} %
\usepackage{multirow} %
\usepackage{color} %
\usepackage{lineno} %
\usepackage{hyperref} %
\usepackage{booktabs}
\usepackage{placeins}
\usepackage{ulem}

\usepackage[usenames,dvipsnames]{xcolor}
\hypersetup{colorlinks=true, urlcolor=black, linkcolor=blue, citecolor=red} %

\usepackage{amsmath}
\usepackage{tcolorbox}
\allowdisplaybreaks[4]

\begin{document}

\title{Scale-invariant enhancement of gravitational waves during inflation}

\author{Atsuhisa Ota}
\email{iasota@ust.hk}
\affiliation{HKUST Jockey Club Institute for Advanced Study, The Hong Kong University of Science and Technology, Clear Water Bay, Hong Kong, P.R.China}

\author{Misao Sasaki}
\affiliation{Kavli Institute for the Physics and Mathematics of the Universe (WPI), University of Tokyo, Chiba, 277-8583, Japan}
\affiliation{Center for Gravitational Physics and Quantum Information, Yukawa Institute for Theoretical Physics, Kyoto University, Kyoto, 606-8502, Japan}
\affiliation{Leung Center for Cosmology and Particle Astrophysics, National Taiwan University, Taipei 10617}
\author{Yi Wang}
\affiliation{Department of Physics, The Hong Kong University of Science and Technology, Clear Water Bay, Hong Kong, P.R.China }
\affiliation{HKUST Jockey Club Institute for Advanced Study, The Hong Kong University of Science and Technology, Clear Water Bay, Hong Kong, P.R.China}

\date{\today}

\begin{abstract}
The inflationary 1-loop tensor power spectrum from an excited spectator scalar field is calculated.
Recent studies on primordial black holes suggest that the inflationary curvature perturbation may be huge on small scales. An enhanced curvature perturbation may arise from a drastic enhancement of spectator scalar field fluctuations. 
In this letter, using the in-in formalism, we calculate 1-loop quantum corrections to primordial gravitational waves by such an excited spectator field with a sharp peak in momentum space.
We find \textit{scale-invariant} loop corrections in this full quantum setup, in contrast to the sharply peaked corrections in the previously calculated scalar-induced tensor modes. Especially on super Hubble scales, the primordial gravitational waves are also amplified, which can be understood as a Bogoliubov transformation of the vacuum due to the excited scalar field.
This mechanism allows us to probe the scalar field properties on extremely short-distance scales with the current and future cosmic microwave background and gravitational wave experiments, opening a novel window for inflationary cosmology.


\end{abstract}

\preprint{YITP-22-84}

\maketitle

As the black hole merger detected in the Laser Interferometer Gravitational-Wave Observatory (LIGO) event in 2015 was unexpectedly massive~\cite{LIGOScientific:2016aoc}, it rejuvenated the idea of primordial black holes~(PBHs)~\cite{Sasaki:2016jop,Clesse:2016vqa,Bird:2016dcv}. A PBH may be formed from collapse of a Hubble horizon size region in the early Universe when the spatial curvature of that region happens to be large and  positive~\cite{Hawking:1971ei,Carr:1974nx,Carr:1975qj}. However the cosmic microwave background~(CMB) data showed tiny curvature fluctuations of $\mathcal O(10^{-5})$ that are almost scale invariant at $k {\rm Mpc/h}\lesssim 0.1$~\cite{Komatsu:2010fb,Planck:2018vyg}.
Hence, the existence of PBHs implies a nontrivial scale dependence of the curvature perturbation at some very short-distance scales, and many models have been proposed to realize such a scale dependence~\cite{Alabidi:2012ex,Drees:2011hb,Drees:2011yz,Garcia-Bellido:2017mdw,Ivanov:1994pa,Ezquiaga:2017fvi,Kannike:2017bxn,Germani:2017bcs,Motohashi:2017kbs,Yokoyama:1998pt,Saito:2008em,Zhou:2020kkf,Chen:2019zza,Pi:2021dft}.

The curvature perturbation, or fluctuations in the energy momentum tensor in general, also produces gravitational waves~(GWs) from nonlinear couplings in the Einstein equation during and/or after inflation (for a review, see e.g.,\cite{Domenech:2021ztg} and references therein). 
As these induced GWs are causally generated, and the spectrum has peaks where the source is amplified, they can be detected in GW experiments as a counter part of PBH formation~\cite{Saito:2008jc,Alabidi:2012ex,Nakama:2016gzw,Inomata:2018epa}. 
We may say, however, that this is a classical effect as the quantum nature does not play any significant role in it.  
Then one may ask a question: how about genuine quantum effects of those fluctuations on cosmological perturbations like curvature and/or tensor perturbations?

\begin{figure}[hbt]
  \includegraphics[width=0.35\textwidth]{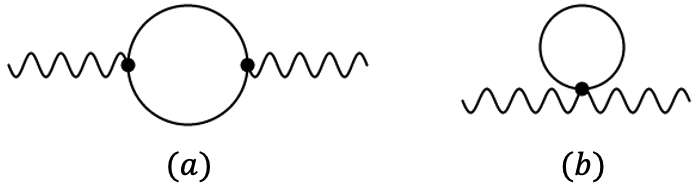}
  \caption{The Feynman diagrams considered here: $(a)$ and $(b)$ corresponds to $P_{h2}$ and $P_{h1}$, respectively, in our calculation.}
  \label{fyn}
\end{figure}

This letter, for the first time, considers the effect of the 1-loop corrections of an excited minimally coupled scalar field to the tensor power spectrum in a full quantum setup.
We focus on the tensor perturbation as it is much simpler than the case of the scalar (curvature) perturbation.
The Feynman diagrams in consideration are shown in Fig.~\ref{fyn}.
We compute those contributions using the in-in formalism.
We find a surprising result that the tensor power spectrum can be scale-invariantly enhanced on superhorizon scales.
This letter concisely reports this novel result. The full technical details will be presented in the companion paper~\cite{Ota:2022xni}.
In our setup, where we are interested in the effect of an excitation, possible divergences in the loop integrals due to infinite large momentum modes or infinitely small momentum modes~\cite{Urakawa:2008rb,Senatore:2009cf,delRio:2018vrj,Tan:2019czo,Comelli:2022ikb,Comelli:2022ikb,Dimastrogiovanni:2022afr} do not appear.
We assume that the regularization and renormalization of the standard ground state cosmological perturbations are applied to remove such possible divergences, which is also assumed for recent induced gravitational wave studies.

Consider a Hamiltonian written as a sum of a free field Hamiltonian and an interaction part, $H_0 + \lambda H_{\rm int}$. 
Then, the vacuum expectation value~(VEV) of a Heisenberg operator $\mathcal O_{\rm H}$ at the conformal time $\tau$ in the interaction vacuum $|\Omega\rangle$ is expressed by the interaction picture field as follows~\cite{Maldacena:2002vr,Weinberg:2005vy}:
\begin{align}
	\langle \Omega |\mathcal O_{\rm H} (\tau) |\Omega\rangle & =  \langle 0 |U(\tau;\tau_0)^\dagger \mathcal O_I(\tau) U(\tau;\tau_0) |0\rangle,\label{eq:vev}
\end{align}
where $|0\rangle$ is the free vacuum, subscript $I$ implies the interaction picture field, and the interaction picture time evolution operator from the initial time $\tau_0$ to $\tau$ is
\begin{align}
	U(\tau;\tau_0)&=T\exp\left(  -i\lambda \int^\tau_{\tau_0}d\tau' H_{{\rm int},I}(\tau') \right),
	\label{uni:def}
\end{align}
with the time ordering operator $T$.
Eq.~\eqref{eq:vev} is expanded into,	 $\langle \Omega |\mathcal O_{\rm H} (\tau) |\Omega\rangle  =\sum_{n=0}\lambda^n \mathcal O_n(\tau),$ where we introduced
\begin{align}
	\mathcal O_0(\tau)&=\langle 0 |\mathcal O_I(\tau)|0\rangle, \label{O:0} 
	\\
	\mathcal O_1(\tau)&= 2 \Im \int^\tau_{\tau_0}d\tau'  \langle 0|\mathcal O_I(\tau)   H_{{\rm int},I}(\tau') |0\rangle,
	\label{O:1}
	\\
	\mathcal O_2(\tau)&=\int^\tau_{\tau_0^*}d\tau'\int^\tau_{\tau_0}d\tau''  
	 \langle 0|  H_{{\rm int},I}(\tau')\mathcal O_I(\tau)   H_{{\rm int},I}(\tau'') |0\rangle
	\notag \\
	-2\Re \int^\tau_{\tau_0}& d\tau' \int^{\tau'}_{\tau_0} d\tau''
	 \langle 0| \mathcal O (\tau)  H_{{\rm int},I}(\tau')H_{{\rm int},I}(\tau'')|0\rangle,
	\label{O:3}
\end{align}
and $\tau_0\equiv -\infty (1-i\epsilon)$ is the initial time of inflation with the infinitesimal rotation on the time contour, to suppress the initial excited states. From now on we will set the order-counting parameter $\lambda$ to $\lambda=1$. 
The Hamiltonian for cosmological perturbations during inflation is obtained by expanding the inflationary full Hamiltonian on a homogeneous and isotropic background spacetime, which is written as
$ 	H_{\rm full} = H^{(0)} + H^{(1)} + H^{(2)}_0 + H^{(>2)}_{\rm int}.$
$H^{(0)}$ is a $c$-number composed of background quantities, and $H^{(1)}$ is the first order term eliminated by the background equation of motion.
$H^{(2)}_0$ is quadratic in perturbation, which accounts for the free theory.
$H^{(>2)}_{\rm int}$ is the rest of the interactions arising from the nonlinearity in the full action,~$S_{\rm full}$.

For the choice of the tensor perturbation variable, we adopt Maldacena's convention~\cite{Maldacena:2002vr}, 
where the spatial component of the spacetime metric is parameterized as
$	g_{ij} = a^2 \left(\delta_{ij}+h_{ij}+\frac{1}{2}h_{i}{}^k h_{kj}+\cdots \right).$
Here $a=-1/(H\tau)$ is the scale factor with the Hubble parameter $H$, and we impose the transverse traceless condition $h^i{}_{i}=\partial_ih^{i}{}_{j}=0$.
Note that we raise and lower the spatial indices of perturbations by regarding them as tensors with respect to the background spatial metric, $\delta^{ij}$ and $\delta_{ij}$.
With this gauge condition, the volume element is unperturbed by $h_{ij}$: $\sqrt{\det |g_{ij}|} = a^3$, so that the interaction between $h_{ij}$ and a minimally coupled scalar field $\chi$ appears only in the kinetic term,
\begin{align}
	S_{\rm full} \supset -\frac{1}{2}\int d^4 x \sqrt{-g} g^{\mu\nu}\partial_\mu  \chi\partial_\nu  \chi.\label{minimalac}
\end{align}
Hereafter we denote the scalar field fluctuation by $\delta  \chi$.
Expanding and Legendre transforming Eq.~\eqref{minimalac}, we find the interaction Hamiltonian for the tensor-scalar coupling 
\begin{align}
		H^{(2>)}_{{\rm int}} \supset &- \frac{1}{2}  \int d^3x  a^2 h^{ij} \partial_i \delta \chi \partial_j \delta \chi
	\notag \\
	&+
	\frac{1}{4}  \int d^3x  a^2h^{ik}h_{k}{}^{j}   \partial_i \delta \chi \partial_j \delta \chi\,,
\label{Hdefint}
\end{align}
where and below we omit the suffix $I$ for the fields in the interaction picture for notational simplicity.
In this work, we do not consider tensor loops but focus on the 1-loop effect due to the amplification of $\delta \chi$.

The perturbation variables in real space are written as 
\begin{align}
		\delta  \chi(\tau, \mathbf x)  &= \int \frac{d^3 q}{ (2\pi)^{3}}e^{i\mathbf q\cdot \mathbf x}\delta  \chi_{\mathbf q}(\tau),
		\label{chi:ft}
		\\	
		h_{ij}(\tau, \mathbf x)  &= \int \frac{d^3 q}{ (2\pi)^{3}}e^{i\mathbf q\cdot \mathbf x} \sum_{s=\pm 2}e_{ij}^s(\hat q)h^s_{\mathbf q}(\tau),
		\label{h:ft}
\end{align}
where the polarization tensors satisfy $e_{ij}^s(\hat q)e^{ij,s'*}(\hat q)  = \delta^{ss'}.$
We recast each term of Eq.~\eqref{Hdefint} as
\begin{align}
	H^{(3)}_{{\rm int}} & =\frac{1}{2}\prod_{A=1}^3\left(\int \frac{d^3p_A}{(2\pi)^{3}}\right)(2\pi)^3\delta\left(\sum_{A=1}^3 \mathbf p_A\right) \sum_{s=\pm2}  
	\notag \\
	\times &  a^2 h^s_{\mathbf p_1} e^{ij,s}(\hat p_1) p_{2i}p_{3j}  \delta \chi_{\mathbf p_2}  \delta \chi_{\mathbf p_3},
\label{H3:def}
\\
		H^{(4)}_{{\rm int}} 
	&=
	 -\frac{1}{4}\prod_{A=1}^4\left(\int \frac{d^3p_A}{(2\pi)^{3}}\right)(2\pi)^3\delta\left(\sum_{A=1}^4 \mathbf p_A\right)  \sum_{s_1,s_2} 
\notag 	 \\
	 \times  a^2  &e^{ik,s_1}(\hat p_1)e_{k}{}^{j,s_2}(\hat p_2) p_{3i}p_{4j}  h^{s_1}_{\mathbf p_1}h^{s_2}_{\mathbf p_2} \delta \chi_{\mathbf p_3}  \delta \chi_{\mathbf p_4}.
\label{H4:def}
\end{align}
The field operators in Fourier space are given by 
	$\delta \chi_{\mathbf q}  =u_q\hat a_{\mathbf q}+ u^*_q\hat a^\dagger_{-\mathbf q}$, and
	$h^s_{\mathbf q}  =v_q\hat b^s_{\mathbf q}+ v^*_q\hat b^{s\dagger}_{-\mathbf q}$,
where $\hat a_{\mathbf q}(\hat b^s_{\mathbf q})$ and $\hat a^\dagger_{-\mathbf q} (\hat b^{s\dagger}_{-\mathbf q})$ are the annihilation and creation operators of the scalar~(tensor) perturbation, respectively, and $u_q$ and $v_q$ are the associated positive frequency mode functions.
For example, the ground state mode functions of scalar and tensor perturbations are written as
	$u_q \equiv  
	H/\sqrt{2q^3}
	(1+iq\tau)e^{-iq\tau}$, and $v_q\equiv 2u_q/M_{\rm pl}$. 
We will compute VEV of $	\mathcal O_I(\tau) = \sum_{s=\pm2}h^s_{\mathbf q}(\tau)h^s_{\mathbf q'}(\tau)$ with these interaction Hamiltonians.

Substituting Eqs.~\eqref{H3:def} and \eqref{H4:def} into \eqref{O:1} and \eqref{O:3}, we find the 1-loop corrections to the tensor power spectrum:
	$P^\text{ 1-loop}_h = P_{h2a}+P_{h2b}+P_{h1},$ 
where we defined
\begin{align}
		P_{h2a}=&   \int_0^{\infty} d p \int_{|p-q|}^{p+q} d\bar p \bar w \left| \int^{\tau}_{\tau_0} d\tau' 
		     a'^2
	v_{q}
	v'^*_{q}
	u'^*_{p}
	u'^*_{\bar p}\right|^2
,	\label{P22a:vev}
\\
		P_{h2b}=&-2 \Re  \int_0^{\infty} d p \int_{|p-q|}^{p+q} d\bar p  \bar w
\notag		\\
	\times &v_q^2
	 \int^{\tau}_{\tau_0} d\tau' a'^2 
	v'^*_{q}
	u'_{p} 
	u'_{\bar p}
	\int^{\tau'}_{\tau_0} d\tau'' 
		   a''^2
	v''^*_{q'}
	u''^*_{p}
	u''^*_{\bar p}  
\label{P22b:vev}
\\
P_{h1}=&   \Im    \int_0^{\infty} dp \frac{p^4}{3\pi^2} 
	\int^{\tau}_{\tau_0} d\tau' 
	  a'^2 v_q^2 v'^{*2}_{q} |u'_{p}|^2\,,	
\label{P13:vev}
\end{align}
where the primes imply those on $\tau$ in the argument, e.g., $a'\equiv a(\tau')$ or $u''_{p}\equiv u_{p}(\tau'')$, and we introduced 
\begin{align}
	\bar w \equiv  \frac{p\bar p \left(p^4-2 p^2 \left(\bar p^2+q^2\right)+\left(\bar p^2-q^2\right)^2\right)^2}{128\pi^2q^5} 
\end{align}
In the derivation of the above equations, bubble graphs and tadpoles are removed.

\begin{figure*}
	\includegraphics[width=\linewidth]{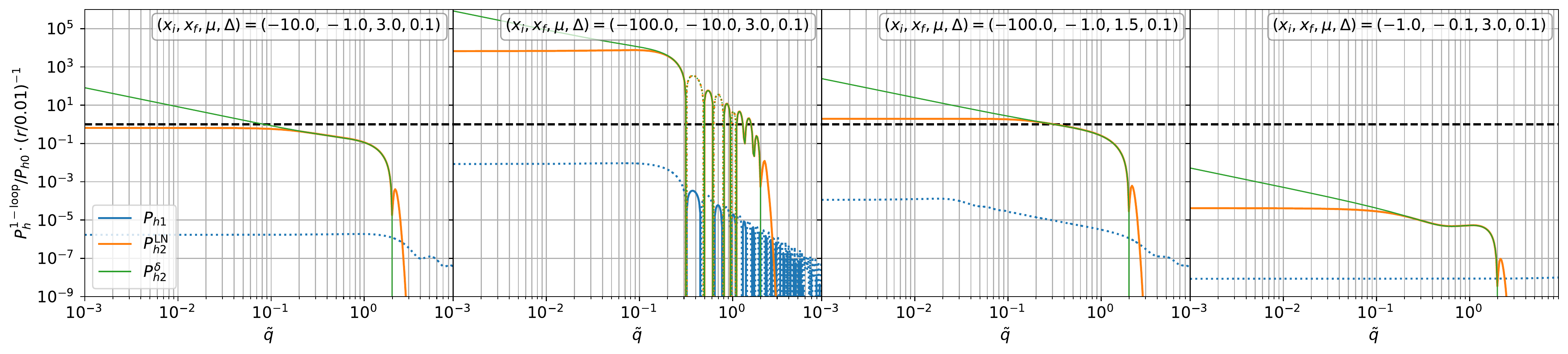}
 	\caption{1-loop spectrum at the end of inflation $\tau=0$ for various parameters associated with the scalar field amplification factors~(see (A) to (D) in the text for details). The horizontal axis is $\tilde q=q/p_*$, the tensor mode's Fourier wavenumber $q$ normalized by the peak location $p_*$. 
 	$x_a=p_*\tau_a$ for $a=i,f$ and $\Delta$ is the width of the log-normal peak.  The orange and blue curves are the loop diagrams $(a)$ and $(b)$ in Fig.~\ref{fyn} for the delta function peak. The narrow log-normal peak counterpart for $(a)$ is illustrated by the thin green curve. Each spectrum is multiplied by $P_{h0}^{-1}\cdot (r/0.01)^{-1}$ with the linear tensor-to-scalar ratio $r$.   The black dashed line indicates unity.
 	Dotted curves represent the negative parts.} 
  	\label{fig2}	
\end{figure*}

For computational simplicity, let us consider that the scalar field fluctuation is amplified only at $p=p_*$ mode as
\begin{align}
	|u_{p}(\tau)|^2 \to \delta(\ln p-\ln p_*)\left(\frac{\tau_i}{\tau }\right)^{2\mu} |u_{p}(\tau)|^2,\label{delta:mode}
\end{align}
which implies that the canonical normalization of the quantum scalar field is changed, for example, by a nontrivial kinetic term.
We consider the resonance starts at $\tau_i\gg \tau_0$ and ends at $\tau_f$.
Eq.~\eqref{delta:mode} is the simplest toy model of excitation, and a different model is also considered in the full paper~\cite{Ota:2022xni}. 
Some models predict the scalar field fluctuation decays after $\tau_f$, but we keep the same value until inflation ends at $\tau \sim 0$ for simplicity.  
The loop integrals are straightforward in this setup.

Using Eq.~\eqref{delta:mode}, Eqs.~\eqref{P22a:vev}, \eqref{P22b:vev}, and \eqref{P13:vev} yield
\begin{align}
	 \frac{P^{\delta}_{h1}}{P_{h0}}=&  \frac{H^2}{M_{\rm pl}^2}   
	 \Im  \int^0_{x_0} dx X_{\tilde q}(x),\label{P13delta}
\\
		\frac{P^{\delta}_{h2a}}{P_{h0}}=& \frac{1}{2}\frac{H^2}{M_{\rm pl}^2} \Theta_{2-\tilde q} 
 \left| \int^{0}_{x_0} Y_{\tilde q} ( x) dx
		     \right|^2,\label{P22adelta}
\\
		\frac{P^{\delta}_{h2b}}{P_{h0}}=&- \frac{H^2}{M_{\rm pl}^2}
		\Theta_{2-\tilde q}
		    \Re \int^{0}_{x_0} dx  \int^{x}_{x_0} dx'  
Y^*_{-\tilde q}(x)Y_{\tilde q}(x'),\label{P22bdelta}
\end{align}	
where the tree level tensor spectrum is $P_{h0} \equiv 2 |v_q(0)|^2$, $x_a\equiv p_*\tau_a$ ($a=0,i,f$), $\tilde q\equiv q/p_*$, $\Theta_{2-\tilde q}$ is the Heaviside step function with the argument $2-\tilde q$ that implies the momentum conservation, and we introduced
\begin{align}
	X_{\tilde q}(x) &\equiv   \frac{(1+ x^2)(1-i\tilde qx )^2}{6\pi^2 \tilde q^3x^{2}}e^{2i\tilde qx} \Xi(x),
	\\
	Y_{\tilde q}(x) &\equiv  \frac{(4-\tilde q^2)(1-i\tilde qx)(1-ix)^2}{16\pi \tilde q^2 x^{2}}
	e^{i(\tilde q+2)x} \Xi(x).
\end{align}
The time dependence of the excited state is represented by $\Xi$, and for the model (\ref{delta:mode}) we have 
$\Xi(x) \equiv (x_i/x_f)^{2\mu}$ for $x_f\leq x$, $(x_i/x)^{2\mu}$ for $x_i\leq x \leq x_f$, and 0 for $x \leq x_i$, where the last condition implies we subtracted the vacuum contribution as commented in the introduction.

In the IR limit $\tilde q\to 0$, we find $P^{\delta}_{h1}/P_{h0}$ is scale invariant, i.e., $P^{\delta}_{h1}$ has the same scaling as $P_{h0}$.
The leading terms of $P^{\delta,{\rm IR}}_{h2a}$ and $P^{\delta,{\rm IR}}_{h2b}$ are $\tilde q^{-4}$, but when combining both terms the exact cancelation of the negative powers happens up to $\tilde q^{-2}$ and we obtain $(P^{\delta}_{h2a}+P^{\delta}_{h2b})/P_{h0}=\mathcal O(\tilde q^{-1})$.
This scaling happens because we considered the delta function spectrum.
The delta function spectrum implies infinite distance correlations in real space, which violates the causality. 
The same issue was discussed in Ref.~\cite{Pi:2020otn} in the context of classically scalar-induced GWs, and they introduced a finite width in the spectrum by considering a log-normal spectrum,
\begin{align}
	\delta(\ln p-\ln p_*) \to \frac{1}{\sqrt{2\pi}\Delta}e^{-\frac{(\ln p/p_*)^2}{2\Delta^2}}.\label{Logintro}
\end{align}
They found an additional power of $\tilde q$ appears from the log-normal factor for the narrow peak~($\Delta \ll 1$) approximation.
The same prescription is applicable to the present case.
Including Eq.~\eqref{Logintro}, the step functions in Eqs.~\eqref{P22adelta} and \eqref{P22bdelta} are generalized to
\begin{align}
&\Theta^\Delta_{2-\tilde q}	 = \frac{e^{2 \Delta ^2}}{2}\left[
\text{erf}\left(\frac{2 \Delta ^2-\ln \left(| 1-\tilde q| \right)}{\sqrt{2} \Delta }\right)
\right.
\notag \\
&\left. -\text{erf}\left(\frac{2 \Delta ^2-\ln \left(1+\tilde q\right)}{\sqrt{2} \Delta }
\right)\right]=\sqrt{\frac{2}{\pi}}\frac{\tilde q}{\Delta} + \mathcal O(\tilde q^3/\Delta^3),
\end{align}
which satisfies $\lim_{\Delta\to 0}\Theta^\Delta_{2-\tilde q}=\Theta_{2-\tilde q}$.
Therefore, in the IR tail, we have $P^{\rm LN,IR}_{h2}=\sqrt{\frac{2}{\pi}}\frac{\tilde q}{\Delta}P^{\delta,{\rm IR}}_{h2}$, so we find scale invariant loop correction at $\tilde q\ll\Delta$.
We also find the contribution of the remaining diagram is unchanged: $P^{\rm LN}_{h1} = P^\delta_{h1}$.

As an example, let us consider $\delta \chi$ is amplified by a factor of $
(x_i/x_f)^\mu \sim 10^3$.
For the same amplification factor, we may consider different situations. 
(A) Near horizon contribution; $x_f=-1$, $x_i=-10$, $\mu=3$: the amplification happens and stops \textit{just before} the horizon exit of a spectator field.
(B) Sub horizon contribution; $x_f=-10$, $x_i=-100$, $\mu=3$: the amplification happens and stops \textit{well in advance} of the horizon exit of a spectator field. 
(C) Sub-to-Near horizon contribution; $x_f=-1$, $x_i=-100$, $\mu=1.5$: the amplification happens \textit{well in advance} and stops \textit{just before} the horizon exit of a spectator field.  
(D) Super horizon contribution; $x_f=-0.1$, $x_i=-1$, $\mu=3$: the amplification happens at the super horizon scale.  
In Fig.~\ref{fig2}, we show the results of numerical calculations for the above (A) to (D).
We present the 1-loop corrections in units of $P_{h0} \cdot (r/0.01)$, where $r$ is the tensor-to-scalar ratio $r$ for linear perturbations.
Hence, the plots above unity mean the loop corrections are bigger than the tree level spectrum for $r=0.01$, and thus the perturbative description may be failed.
When $\delta \chi$ is on the super horizon scale in (D), the loop corrections are relatively suppressed because of causality but are not exactly zero.
The figure shows that the amplification at a shorter scale introduces larger 1-loop corrections for the same amplification factor.
This is because the amplified $\delta \chi$ continuously contributes before the horizon exit.
The IR behavior discussed analytically is reproduced in numerical calculation.
The size of $P_{h1}$ and $P_{h2}$ are loosely related as $P_{h2}\approx (x_i/x_f)^{2\mu} P_{h1}$ near $\tilde q=1$ as the former involves two additional scalar field operators.

One might be interested in the connection to the induced tensor modes that we mentioned earlier.
The generation of the scalar-induced tensor mode is a classical process in the sense that we integrate the classical equation of motion.
A peculiar solution in the presence of a classical source is~\cite{Domenech:2021ztg}
\begin{align}
	h^{s,{\rm ind.}}_{\mathbf q} = \int d\tau' G_{\mathbf q}(\tau,\tau')S^s_{\mathbf q}(\tau'),\label{ind}
\end{align}
where $G_{\mathbf q}(\tau,\tau')$ is the retarded Green function and $S^s_{{\rm cl.},\mathbf q}$ is the source term for the polarization $s$, which arises from the first line of the interaction Hamiltonian (\ref{Hdefint}), or $H^{(3)}_{\rm int}$ in Eq.~(\ref{H3:def}). In the quantum language, this corresponds to computing the first order effect of the interaction Hamiltonian $H^{(3)}_{\rm int}$.

Contrary to conventional wisdom, we find that tensor modes can evolve after the horizon exit. 
Physically, the interaction with the excited scalar field redefines the vacuum state and hence modifies the mode functions.
As discussed below, the quantum evolution of the super horizon mode may be interpreted as a Bogoliubov transformation, which is not prohibited from causality.

As we are employing the interaction picture where quantum properties play the essential role, it is expected that if we can solve the evolution of the Heisenberg operator perturbatively but explicitly, we should be able to clearly see how the quantum nature comes into play. 
Using $\mathcal O_{\rm H}(\tau) = U(\tau;\tau_0)^\dagger \mathcal O_I(\tau) U(\tau;\tau_0)$ for the Heisenberg operator $\mathcal O_{\rm H}= h^s_{\mathbf q,{\rm H}}$ with Eq.~\eqref{uni:def}, up to second order, one finds~\cite{Weinberg:2005vy}
\begin{align}
	&h^s_{\mathbf q,{\rm H}}(\tau) =h^s_{\mathbf q}(\tau) +  i \int^\tau_{\tau_0} d\tau' [H_{{\rm int}}(\tau'),h^s_{\mathbf q}(\tau)]\notag \\
&-\int^\tau_{\tau_0} d\tau' \int^{\tau'}_{\tau_0} d\tau'' [H_{{\rm int}}(\tau''),[H_{{\rm int}}(\tau'),h^s_{\mathbf q}(\tau)]]\,,\label{h_heisenberg}
\end{align}
where the first order interaction Hamiltonian describes the Born approximation, while the second order term is the first iterative contribution.
One clearly sees that Eq.~\eqref{ind} in quantum theory is written as~\cite{Fumagalli:2021mpc}  
\begin{align}
	h^{s,{\rm ind.}}_{\mathbf q} = i \int^\tau_{\tau_0} d\tau' [H^{(3)}_{{\rm int}}(\tau'),h^s_{\mathbf q}(\tau)].\label{ind_heisenberg}
\end{align}
The remaining terms are not necessarily specific to quantum theory but have been ignored in the previous works. One of the authors recently considered the classical counterpart of those contributions in Ref.~\cite{Chen:2022dah} and found similar one-loop order super horizon variation.
We are interested in how the initial vacuum state $h^s_{\mathbf q}$ evolves to a final state. Therefore, to find the relevant part of $h^s_{\mathbf q.{\rm H}}$, 
we compute the remaining Born approximation part due to $H_{\rm int}^{(4)}$ and the iterative contribution due to $H^{(3)}_{\rm int}$.
After some algebra and taking the average over $\delta\chi$, we find
\begin{align}
	\langle h^s_{\mathbf q,{\rm H}}\rangle_{\delta \chi} =V_{q}b^s_{\mathbf q} + V^*_{q} b^{s\dagger}_{\mathbf q},~V_{q} = \alpha_q v_q +\beta_q v_q^*,
	\label{bogo_eq}
\end{align}
where $\langle \cdots\rangle_{\delta \chi}$ implies that we integrated out $\delta \chi$, and $\alpha_q$ and $\beta_q$ are functions of $\tau$ due to the interaction that are found to satisfy $|\alpha_q|^2-|\beta_q|^2=1$ up to 1-loop order. Thus, we may regard the linear transformation from $(v_q,v_q^*)$ to $(V_q,V_q^*)$ as the Bogoliubov transformation.
Those details will be presented in the companion paper~\cite{Ota:2022xni}.
In this regard, it is interesting to note that the diagram $(b)$ in Fig.~\ref{fyn}, which is due to the 
Born approximation part of $H^{(4)}_{\rm int}$ and the iterative contribution,
may be regarded as an effective time-dependent mass term for the tensor perturbation. In a class of massive gravity theory, it is known that such a term can enhance the tensor perturbation on superhorizon scales \cite{Lin:2015nda}. A similar term also arises in the present case, as our background has no time-translational invariance.

To justify our perturbative analysis, the scalar field's energy density should be, at least, sufficiently smaller than the background energy density during inflation: $\langle a^{-2} (\partial \delta \chi)^2\rangle \ll M_{\rm pl}^2H^2$, which can be recast into $(p_*/a)^2(\tau_i/\tau)^{2\mu}\ll M_{\rm pl}^2$.
In the above example, we considered $(\tau_i/\tau)^{\mu}\lesssim 10^3$, which gives the condition for the peak physical momentum: $p_*/a\ll 10^{15}{\rm GeV}$. This condition may be easily satisfied for the scale of our interest.

In this letter, we only considered the 1-loop corrections to the tensor power spectrum. But the 1-loop tensor bispectrum involves more scalar propagators as it contains six powers of $\delta\chi$, $\mathcal O(\delta \chi^6)$. Hence, it will also be amplified. Thus, the tensor perturbation may become highly non-Gaussian due to quantum corrections.
We note that the 1-loop corrections to the scalar power spectrum give rise to a similar structure.
The difference is that the interaction terms are higher-order in slow-roll parameters~\cite{Maldacena:2002vr}, which could be relatively suppressed compared to the present case.
We leave a detailed analysis of the scalar case to a future project.
Note that for some non-attractor models~\cite{Garcia-Bellido:2017mdw} where large curvature perturbations are produced by modifying the slow-roll parameters, the loop corrections will not be amplified because the scalar-tensor coupling in Eq.~\eqref{Hdefint} is slow-roll suppressed.

For the time integration, we included the range $x_f<x<0$, where the scalar fluctuations are no longer enhanced but assumed to remain constant. This results in the change in the sign of $d^2 P_h/d\tau^2$ at $x=x_f$ in our numerical calculation.
The late time contribution is found to be a factor of $\mathcal O(1)$ to $\mathcal O(10)$ of the early time contribution, depending on the parameters, but the IR scaling does not change. 
In general, the field fluctuations may decay after the enhancement, depending on the models. In such a case, there will be virtually no appreciable late time contribution. In any case, as the final amplitude of the tensor spectrum depends largely on the late time evolution of the scalar field fluctuations, precise specifications of them are necessary to make quantitatively accurate predictions. 

Finally, we mention a potential issue about the gauge independence. The gauge (in)dependence of $P_{h1}$ is not obvious.
In the Maldacena gauge, which we adopted, the fourth-order coupling appears only in the scalar field kinetic term for a minimally coupled scalar field.
In other gauges, a non-vanishing contribution from $\sqrt{-g}$ introduces additional gauge-dependent terms. It is interesting to study if the gauge independence of $P_{h1}$ can be shown when the contributions of all these gauge-dependent terms are combined.

To conclude, we found that the quantum corrections to the tensor spectrum can be scale-invariant and large; hence it will have a strong impact on GW measurements at all scales~\cite{Kogut:2011xw,Andre:2013afa,Matsumura:2013aja,Desvignes:2016yex,Brazier:2019mmu,Kerr:2020qdo,LISA:2017pwj,Kawamura:2011zz,Ruan:2018tsw,TianQin:2015yph}, in contrast to the induced GWs that could be measurable only if the scalar spectrum {\it by happy chance} has peaks at scales relevant to the current and future GW detectors.

\begin{acknowledgments}
The authors would like to thank Yifu Cai, Ema Dimastrogiovanni, Guillem Domenech, Keisuke Inomata, and Jiro Soda for their useful comments on a preliminary version of the paper.
AO and YW are supported in part by the National Key R\&D Program of China (2021YFC2203100), the NSFC Excellent Young Scientist Scheme (Hong Kong and Macau) Grant No.~12022516, and by the RGC of Hong Kong SAR, China (CRF C6017-20GF and GRF 16303621).
MS is supported in part by the JSPS KAKENHI grant Nos.~19H01895, 20H04727, and 20H05853.
\end{acknowledgments}
\appendix

\bibliography{sample.bib}{}
\bibliographystyle{unsrturl}

\end{document}